\input{aipcheck}

\documentclass[
    ,draft            
    ,numberedheadings 
  ]
  {aipproc}

\layoutstyle{6x9}

\begin{document}

\title{The directed-loop algorithm}

\author{Anders W. Sandvik}{
 address={Department of Physics, {\AA}bo Akademi University, 
Porthansgatan 3, FIN-20500 Turku, Finland}
}

\author{Olav F. Sylju{\aa}sen}{
address={NORDITA, Blegdamsvej 17, DK-2100 Copenhagen {\O}, Denmark}
}

\begin{abstract}
The directed-loop scheme is a framework for generalized loop-type updates in 
quantum Monte Carlo, applicable both to world-line and stochastic series
expansion methods. Here, the directed-loop equations, the solution of which
gives the probabilities of the various loop-building steps, are 
discussed in the context of the anisotropic $S=1/2$ Heisenberg model in 
a uniform magnetic field. This example shows how the directed-loop concept
emerges as a natural generalization of the conventional loop algorithm, 
where the loops are selfavoiding, to cases where selfintersection must be 
allowed in order to satisfy detailed balance.
\end{abstract}

\maketitle


\section{Introduction}

Loop algorithms \cite{evertz1,kawashima1,evertz2} have dramatically improved 
the performance of world-line quantum Monte Carlo calculations 
\cite{worldline}. The autocorrelation times can be reduced by several orders 
of magnitude relative to standard local updating schemes \cite{kawashima2}. 
However, the conventional loop updates are restricted to certain models and/or
limited regions of their parameter spaces. In particular, external fields 
cannot be taken into account when constructing a loop, and the loop-flip is 
then conditional upon a subsequent Metropolis \cite{metropolis} accept/reject 
step. The acceptance probability for large loops in a high field is small, 
and this approach is therefore feasible only at high temperatures or very 
weak fields \cite{troyer1}. The restriction is analogous to that in classical 
Monte Carlo, where cluster 
algorithms \cite{sw,wolff} also are not applicable to spin models in a 
magnetic field. Remarkably, two recent generalizations of the loop concept 
have overcome this problem for quantum systems. The worm algorithm \cite{worm}
for world-lines in continuous imaginary time and the operator-loop algorithm 
\cite{sseloop} for stochastic series expansion (SSE) \cite{sse} generalize 
the loop by allowing it to selfintersect and backtrack. The original 
prerequisite of a cluster algorithm, i.e., to express the partition function 
using new auxiliary variables \cite{sw,fk}, is then circumvented, and the 
generalized loops can therefore take complicated interactions and external 
fields into account. The loop-building takes place in an extended 
configuration space of the original variables (spin states or occupation 
numbers), where configurations with uncompleted loops (or worms) do not 
contribute to the partition function (they correspond to violation of a 
conservation law). Such a method was attempted already in the early days of 
the world-line algorithm \cite{cullen}, but without enforcing detailed 
balance in the loop construction. Due to the low acceptance probability for 
random-walk loops, the method was not as efficient as simple local updating 
schemes \cite{worldline,ding}. In the worm and SSE operator-loop algorithms, 
detailed balance is ensured by local probabilistic rules, and the resulting 
closed-loop configurations are always accepted.

The exact relationship between the conventional loop algorithms 
\cite{evertz1,kawashima1,kawashima2} and the more general loop-type algorithms 
allowing selfintersection and backtracking \cite{worm,sseloop} was not 
immediately clear. In particular, the general algorithms did not reduce 
to a standard loop algorithm in regions of parameter space where such an 
algorithm could be applied. Although the efficiency was dramatically improved 
over local updates \cite{kashurnikov,sseloop,dorneich,hebert}, the standard 
loop algorithm was still much more efficient when applicable. Particularly 
unsatisfying was the fact that simulations could not be carried out as 
effectively close to a region of an applicable conventional
loop algorithm as within such 
a region. This ``algorithmic discontinuity'' problem was solved with the 
introduction of the {\it directed-loop algorithm} \cite{dloop}, 
which can often be tuned so that the probabilities of 
selfintersection and backtracking smoothly vanish as a region of an applicable
loop algorithm is approached. The directed-loop algorithm then becomes 
identical to a standard single-loop algorithm (i.e., one loop at 
a time is constructed, as in the classical Wolff cluster algorithm 
\cite{wolff}). The directed loops thus emerge as a natural generalization 
of the original \cite{evertz1} loop concept, in a way similar to the 
Kandel-Domany generalization \cite{kd} of classical cluster 
algorithms.

In the directed-loop scheme, the detailed-balance conditions lead to a set 
of coupled equations for the probabilities of the various loop-building 
(or worm\footnote{Generally speaking, a worm is a different name for an 
incomplete loop, but it should be noted that the worm-building processes 
in the worm algorithm \cite{worm} differ from those used in SSE operator
loops \cite{sseloop,dloop} and the directed loops for world-lines in 
continuous or discrete imaginary time \cite{dloop}.}) steps. 
These {\it directed-loop equations} often 
have an infinite number of solutions, which hence should be optimized. The 
directed-loop algorithm was first developed 
for SSE, but adaptations to world-lines, both in discrete and continuous 
imaginary time, were also presented in the same article \cite{dloop}. 
Conceptually, the scheme is simpler (and often more efficient)
for SSE, and here it will therefore be discussed only within this
representation. For simplicity, only the $S=1/2$ XXZ model (the anisotropic 
Heisenberg model in a uniform magnetic field) will be considered. In this case 
the optimization criterion for the directed loops is taken to be the 
minimization of the backtracking probability. 

In Sec.~2 the basics of the SSE method are reviewed, first in general and
then focusing on the details for the $S=1/2$ XXZ model. The structure of
the operator loops and the derivation of the directed-loop equations are 
discussed in Sec.~3. Some recent applications and extensions of the
directed-loop algorithm are summarized in Sec.~4.

\section{Stochastic Series Expansion}

The SSE method \cite{sse} is an efficient and widely applicable generalization
of Handscomb's \cite{handscomb} power-series method. To construct the SSE 
representation of the partition function, 
$Z={\rm Tr }\{{\rm exp}(-\beta H) \}$, 
the Hamiltonian is first written as a sum,
\begin{equation}
H = - \sum\limits_{a}\sum\limits_{b} H_{a,b},
\label{hsum}
\end{equation}
where in a chosen basis $\{ |\alpha \rangle \}$ the operators satisfy
$H_{a,b}|\alpha \rangle \sim |\alpha^\prime \rangle$, where $|\alpha \rangle$ 
and $|\alpha^\prime \rangle$ are both basis states. The subscripts $a$ and $b$
refer to the operator types (various diagonal and off-diagonal terms) and the
lattice units over which the interactions are summed (e.g., the bonds 
corresponding to two-body interactions). A unit operator  $H_{0,0} \equiv 1$ 
is also defined. Using the Taylor expansion of exp$(-\beta H)$ truncated at 
order $M$, the partition function can then be written as \cite{sse}
\vskip-2mm
\begin{equation}
Z = \sum\limits_\alpha \sum_{S_M} {\beta^n(M-n)! \over M!} 
    \left \langle \alpha  \left | \prod_{p=1}^M H_{a_p,b_p} 
    \right | \alpha \right \rangle ,
\label{zm}
\end{equation}
where $S_M = [a_1,b_1],[a_2,b_2],\ldots ,[a_M,b_M]$ corresponds to the 
operator product, and $n$ denotes the number of non-$[0,0]$ elements (i.e., 
the actual expansion-order of the terms). $M$ can be adjusted during the 
equilibration of the simulation, so that it always exceeds the highest power 
$n$ reached; $M \to An_{\rm max}$, where, e.g., $A = 4/3$. 
Then $M \sim \beta N$, where $N$ is the number of sites, and the remaining 
truncation error is completely negligible. 
Defining a normalized state $|\alpha (p)\rangle$ as $|\alpha \rangle$ 
propagated by the first $p$ operators,
\vskip-2.5mm
\begin{equation}
|\alpha (p)\rangle \sim \prod_{i=1}^p H_{a_i,b_i} |\alpha \rangle,
\label{propagated}
\end{equation}
the periodicity $|\alpha (M) \rangle = |\alpha (0) \rangle$ is required for 
a non-zero contribution to $Z$. In an SSE simulation, transitions 
$(\alpha,S_M) \to (\alpha^\prime, S_M^\prime)$ satisfying detailed balance 
are carried out to sample the configurations. Three classes of updates are 
typically used:

(i) {\it Diagonal update}, where the expansion order $n$ is changed by 
replacing a fill-in unit operator by a diagonal operator from the sum 
(\ref{hsum}), or vice versa, i.e., $[0,0] \leftrightarrow [d,b]$, where 
the type-index $d$ corresponds to a diagonal operator in the basis used. 

(ii) {\it Off-diagonal update}, where a set of operators $\{ [a_p,b_p]\}$
is updated by changing only the type-indices $a_p$. Off-diagonal operators 
cannot be added and removed one-by-one with the periodicity constraint 
$|\alpha (M)\rangle = |\alpha (0)\rangle$ maintained. Local updates involving 
two simultaneously replaced operators can be used \cite{sse}, but much more 
efficient loop \cite{sseloop} and ``quantum-cluster'' \cite{sseising} 
updates have also been developed. 
 
(iii) {\it State update}, which affects only the state $|\alpha \rangle$
in (\ref{zm}). 
This state is just one out of the whole cycle of propagated states 
$|\alpha (p)\rangle$, and it can change in the off-diagonal updates (ii).
However, at high temperatures many sites will frequently have no operators 
acting on them, and they will then not be affected by off-diagonal updates. 
The states at these sites can then instead be randomly modified, as they do 
not affect the weight. 

Turning now to the anisotropic Heisenberg antiferromagnet in a 
magnetic field,
\begin{equation}
H = J \sum_{\langle i,j\rangle}[S^x_iS^x_j + S^y_iS^y_j +
\Delta S^z_iS^z_j] - h \sum_{i}S^z_i,~~~~(J>0, \Delta \ge 0),
\label{xxzmodel}
\end{equation}
the standard $z$-component basis is used:
$|\alpha \rangle = |S^z_i,\ldots,S^z_N\rangle,~~ S^z_i = \pm 1/2$.
Diagonal and off-diagonal bond operators are defined,
\begin{eqnarray}
H_{1,b} & = & \epsilon + {\Delta/4} + h_b 
- \Delta S^z_{i(b)}S^z_{j(b)} + h_b[S^z_{i(b)}+S^z_{j(b)}], \label {h1b} \\
H_{2,b} & = & -\hbox{$1\over 2$}[S^+_{i(b)}S^-_{j(b)} + S^-_{i(b)}S^+_{j(b)}],
\label{h2b}
\end{eqnarray}
where $i(b),j(b)$ are the sites connected by bond $b$ and $h_b$ is the
bond-field (e.g., on a $d$-dimensional cubic lattice $h_b=h/2d$). 
The Hamiltonian can now be written in the form 
(\ref{hsum}) with $a=1,2$ and $b=1,\ldots N_b$, where $N_b$ is the number 
of bonds (e.g., $N_b=dN$ on a $d$-dimensional cubic lattice). Note again
that the unit operator $H_{0,0}=I$ is not part of the Hamiltonian; it is
a fill-in element for augmenting the products of order $n < M$ 
in (\ref{zm}).

The constant $\epsilon + {\Delta\over 4} + h_b$ has been added to the diagonal
bond-operator (\ref{h1b}) in order to render all its matrix elements positive
($\epsilon \ge 0$). On a bipartite lattice, the minus-sign in the off-diagonal
operator (\ref{h2b}) is irrelevant, and the expansion (\ref{zm}) is then 
positive-definite. The sign problem for frustrated $XY$-interactions 
\cite{henelius} will not be considered here.

Storing the operator sequence $S_M$ and a single state $|\alpha (p)\rangle$
(initially $|\alpha \rangle = |\alpha (0)\rangle$),
diagonal updates of the form $[0,0] \leftrightarrow [1,b]$ can be carried 
out sequentially for $p=1,\ldots,M$ at all elements $[a_p,b_p]$ in $S_M$ 
with $a_p =0,1$. The Metropolis acceptance probabilities for such 
substitutions are \cite{sse}
\begin{eqnarray}
P([0,0] \rightarrow [1,b]) & = & 
N_b\beta \langle S^z_{i}(p)S^z_{j}(p)|H_{1,b}|
S^z_{i}(p)S^z_{j}(p)\rangle /(M-n), \label{padd} \\
P([1,b] \rightarrow [0,0]) & = & 
(M-n+1)/ [N_b\beta \langle S^z_{i}(p)S^z_{j}(p)|H_{1,b}|
S^z_{i}(p)S^z_{j}(p)\rangle ],\label{pdel} 
\end{eqnarray}
where, as always \cite{metropolis}, $P >1$ should be interpreted
as probability one. The spins $S^z_i(p)$ refer to the propagated states 
(\ref{propagated}), which are generated one-by one during the diagonal update
by flipping spins whenever off-diagonal operators $[2,b]$ are encountered. 

In the early applications of the SSE scheme \cite{sse}, local off-diagonal
updates involving simultaneous substitution of two operators were used, i.e.,
$[1,b_{p}][1,b_{q}] \leftrightarrow [2,b_{p}][2,b_{q}]$. The
operator-loop update \cite{sseloop} to be discussed next is a much more 
efficient way of sampling the off-diagonal operators.

\section{Operator loops and Directed loops}

\begin{figure}
\includegraphics[height=5.8cm]{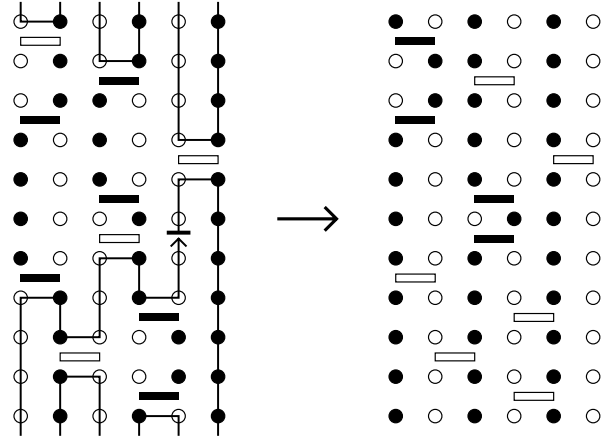}
\caption{An SSE configuration of order $n=10$ for a 6-site isotropic XXZ 
chain. Open and solid circles correspond to up and down spins, and open and 
solid bars represent diagonal and off-diagonal operators, respectively. The 
construction of a loop is illustrated to the left, where the start/end point 
is indicated with a bar/arrow, and he initial direction of movement is 
upward. The configuration obtained when the loop has been flipped is 
shown to the right.}
\label{sseloop}
\end{figure}

To begin the discussion of SSE loop-type updates, it is useful to first 
consider one of the simplest cases; the isotropic model, with $\Delta=1, h=0$
in (\ref{xxzmodel}). In this case, setting $\epsilon=0$ in Eq.~(\ref{h1b}), 
both the diagonal ($[1,b]$) and off-diagonal ($[2,b]$) operators can act only 
on anti-parallel spins, and the corresponding matrix elements are $1/2$ 
[neglecting the negative sign in (\ref{h2b})]. Fig.~\ref{sseloop} shows a 
graphical representation of a valid configuration, along with an illustration 
of a {\it deterministic loop update} \cite{sseloop}. Here all fill-in 
operators $[0,0]$ and the corresponding propagated states have been left out 
since they are irrelevant in the loop update (as the expansion-order $n$ does 
not change). Selecting the starting point and a direction (up or down) at 
random, a loop is constructed using a completely deterministic rule: Moving 
along the chosen direction, whenever an operator is encountered the path 
switches to the other spin connected to that operator, and the direction of 
movement is reversed. This will eventually lead to a closed loop when the 
initial starting point is reached. A new valid configuration is then obtained 
by flipping the spins along the loop and changing the types of all operators 
encountered; diagonal $\leftrightarrow$ off-diagonal (in practice, the changes
are carried out on the run while building the loop). Operators encountered 
twice will remain unchanged. Since all non-zero matrix elements of the bond 
operators equal $1/2$, the new configuration has exactly the same weight as 
the old one, and the loop-flip can hence always be accepted. This type of 
loop is self-avoiding by construction. Instead of constructing loops 
one-by-one and flipping them with probability one, the configuration can 
therefore also be decomposed into all its loops (each spin belongs to
exactly one loop), which are flipped independently of each other with 
probability $1/2$ (as in the classical Swendesen-Wang \cite{sw} algorithm).

The deterministic loop update clearly relies on the symmetry of the model 
at the isotropic point ($\Delta=1,h=0$). The simple rule of ``switch and 
reverse'' when an operator is encountered has to be modified in order to 
construct a more general loop-scheme, applicable for any $\Delta, h$. In 
the general operator-loop update \cite{sseloop}, there are four possibilities
for the worm-like path to proceed when an operator is 
encountered; it can continue on the same spin or switch to the other spin 
connected by the operator, and in either case the direction of the movement 
can be up or down. The directed-loop approach provides the general 
detailed-balance conditions that these probabilities have to satisfy.

\begin{figure}
\includegraphics[height=3.3cm]{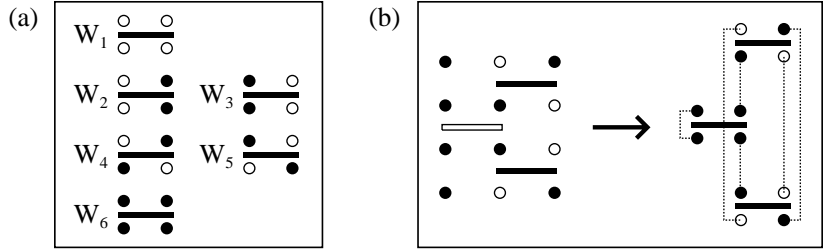}
\caption{(a) All vertices for the $S=1/2$ XXZ model, with their vertex 
weights $W_k$. (b) The linked-vertex representation (right) of a full
3-spin SSE configuration with $n=3$ (left).}
\label{vertices}
\end{figure}

In order to discuss the general operator-loop update and the directed-loop 
scheme, it is useful to introduce a different representation of the SSE
configurations. It is not necessary to store the full states 
$|\alpha (p)\rangle$ shown in Fig.~\ref{sseloop}; the same-spin ``lines'' 
between the operators clearly contain a great deal of redundant information. 
One can represent the matrix element in Eq.~(\ref{zm}) as a linked lists 
of {\it vertices} \cite{sseloop}. Note first that the weight of a 
configuration $(\alpha,S_M)$ can be written as
\begin{equation}
W(\alpha,S_M) = {\beta^n(M-n)! \over M!} \prod_{p=1}^n  W(p),
\label{wasm}
\end{equation}
where the product is over the $n$ non-$[0,0]$ operators in $S_M$.
$W(p)$ will be referred to as a {\it bare vertex weight}; it can be written
as a matrix element of the full bond operator 
$H_b = H_{1,b}+H_{2,b}$ at 
position $p$;
\begin{equation}
W(p) = \langle S^z_{i(b_p)}(p)S^z_{j(b_p)}(p) |H_{b_p}|
S^z_{i(b_p)}(p-1)S^z_{j(b_p)}(p-1)\rangle .
\label{wp}
\end{equation}
A vertex represents the spins on bond $b_p$ 
before and after the operator has acted. These four 
spins constitute the {\it legs} of the vertex. There are six allowed 
vertices, with four different vertex-weights as illustrated in 
Fig.~\ref{vertices}(a). The weights are
\begin{eqnarray}
W_1 &=& \langle \downarrow \downarrow | H_b | \downarrow \downarrow \rangle 
~~=~~ \epsilon, \nonumber \\
W_2 &=& \langle \downarrow \uparrow | H_b | \downarrow \uparrow \rangle~~=~~
W_3 ~~=~~ \langle \uparrow \downarrow | H_b | \uparrow \downarrow \rangle~~=~~
\Delta /2 + h_b + \epsilon,  \label{bweights}  \\
W_4 &=& \langle \uparrow \downarrow | H_b | \downarrow \uparrow \rangle~~=~~
W_5 ~~=~~ \langle \downarrow \uparrow | H_b | \uparrow \downarrow \rangle~~=~~ 
1/2, \nonumber \\
W_6 &=& \langle \uparrow \uparrow | H_b | \uparrow \uparrow \rangle~~=~~
\epsilon + 2h_b.   \nonumber  
\nonumber
\end{eqnarray}
An example of a linked-vertex representation of a term with three bond
operators is shown in Fig.~\ref{vertices}(b). The links connect vertex-legs 
on the same site, so that from each leg of each vertex, one can reach the 
next or previous vertex-leg on the same site (i.e., the links are 
bidirectional). In cases where there is only one operator acting on a given 
site, the corresponding ``before'' and ``after'' legs of the same vertex 
are linked to each other [as is the case with the legs on site 1 in 
Fig.~\ref{vertices}(b)]. 

The building of a loop in the linked-vertex representation consists of a series
of steps, in each of which a vertex is entered at one leg (the entrance leg) 
and an exit leg is chosen according to probabilities that depend on the 
entrance leg and the spin states at all the legs [i.e., the vertex type, 
$k=1,\ldots ,6$, in Fig.~\ref{vertices}(a)]. The entrance to the following 
vertex is given by the link from the chosen exit leg. The spins at all visited
legs are flipped, except in the case of a {\it bounce}, where the exit is
the same as the entrance leg, and only the direction of movement is
reversed. The starting point of the loop is chosen at random. Two 
{\it link-discontinuities} (which are analogous to the source operators in the 
worm algorithm \cite{worm}) are then created when the first entrance 
and exit spins are flipped, i.e., these legs will now be linked to legs 
with different spins . Configurations contributing to $Z$ only contain 
links between  same-spin legs [as in Fig.~\ref{vertices}(b)]. When the 
loop closes, the two discontinuities annihilate each other, and a new 
contributing configuration has then been generated.

The probabilities for the different exit legs 
($e=1,\ldots 4$), given the type of the
vertex ($k=1,\ldots 6$) and an entrance leg ($i=1,\ldots 4$), are 
chosen such that detailed balance is satisfied. This leads to the 
directed-loop equations, 
which are constructed in the following way: Unknown weights $a_e(i,k)$
are first assigned to all possible paths ($i \to e$) through each vertex $k$. 
The sum of all these path weights over all exits $e$ must equal the 
bare vertex weight (\ref{wp}),
i.e., the matrix element before the entrance and exits spins have been 
flipped; $\sum_e a_e(i,k)=W_k$. The actual normalized exit probability is the 
path weight divided by the bare vertex weight; $P_e(i,k)=a_e(i,k)/W_k$. The 
key observation leading to the directed-loop equations \cite{dloop} is that 
the weights for vertex-paths $i \to e$ that constitute each other's reverses 
have to be equal: If the path $i\to e$ through vertex $k$ leads to the vertex
$k'$ when the entrance and exit spins have been flipped, then the reverse path 
$e \to i$ through $k'$ yields vertex $k$, and if $a_e(i,k)=a_i(e,k')$ it
immediately follows that $P_e(i,k)W_k=P_i(e,k')W_{k'}$, i.e., local detailed
balance is satisfied. 

\begin{figure}
\includegraphics[height=2.8cm]{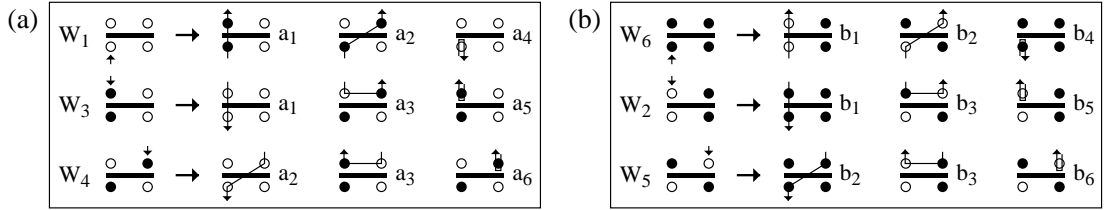}
\caption{Two closed sets of vertex paths, with their corresponding bare
vertex weights $W_k$ and path weights $a_j$, $b_j$. The entrance legs are
indicated with arrows pointing into the bare vertices, and the exit legs
for the three allowed paths are at the arrows pointing out from the
vertices. The entrance and exit spins on the paths have been flipped.}
\label{dpaths}
\end{figure}

The condition $a_e(i,k)=a_i(e,k')$ couples some of the equations 
$\sum_e a_e(i,k)=W_k$ for different entrance legs $i$ and vertices $k$, 
and these equations have to be solved for all the weights 
$a_e(i,k)$. Typically, not all the equations are coupled, however,
but there are several different sets that can be solved independently of 
each other. Such {\it closed sets} for the XXZ model are illustrated in 
Fig.~\ref{dpaths}. Here the path-weights are labeled $a_i$ and $b_i$ for 
the two sets (a),(b), and paths that constitute each other's reverses 
have been assigned the same weight. Note that one of the four exits 
always leads to a new vertex that does not correspond to a term in 
the XXZ Hamiltonian; these paths are not allowed and are not included in 
the figure. The directed-loop equations for the two closed sets are 
\begin{eqnarray}
&& W_1 = a_1 + a_2 + a_4,~~~~~~  W_6 = b_1 + b_2 + b_4, \nonumber \\
&& W_3 = a_1 + a_3 + a_5,~~~~~~  W_2 = b_1 + b_3 + b_5, \label{eqsets} \\
&& W_4 = a_2 + a_3 + a_6,~~~~~~  W_5 = b_2 + b_3 + b_6, \nonumber
\end{eqnarray}
where the bare vertex weights $W_k$ are given in Eq.~(\ref{bweights}).
All the remaining closed sets are related to those in Fig.~\ref{dpaths} by 
trivial symmetries, and for $h=0$ the corresponding two sets of equations 
(\ref{eqsets}) are identical. Note that there are six weights $a_i$ and 
$b_i$ to be solved for in each set, 
but only three equations. There is thus an infinite
number of solutions, even with the requirement that all weights have to be 
positive (since the probabilities are obtained by dividing by the 
positive matrix elements $W_k$). 

\begin{figure}
\includegraphics[height=4.25cm]{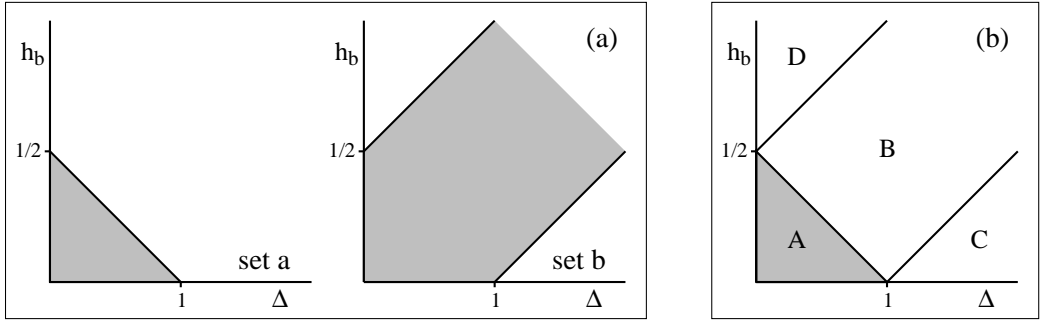}
\caption{(a) The shaded areas show the regions of the $\Delta\ge 0,h\ge 0$
parameter space where the solutions of the directed-loop equations are 
bounce-free (set $a$ left, set $b$ right). (b) Algorithmic phase diagram.
The labels A,B,C,D correspond to the solutions for the weights listed
in Table.~1.}
\label{nobounce}
\end{figure}

In Ref.~\cite{sseloop}, a particular
``heat-bath'' solution was obtained by working directly with the probabilities,
instead of analyzing the path-weights of the directed-loop scheme. 
The directed-loop equations (\ref{eqsets}) provide a more
general framework for finding the optimal solution, i.e., the one which 
leads to simulations with the shortest  autocorrelation times.
There is currently no rigorous way of finding the optimal solution, but
heuristic arguments have been put forward \cite{dloop}: It is a reasonable 
assumption that the probabilities for the bounce processes (i.e., the last
columns in the sets in Fig.~\ref{dpaths}) should be minimized, as they do 
not accomplish any vertex changes and cause the loop-building process to 
backtrack one step (and sometimes more than one step as the loop-building 
continues in the opposite direction). For the model considered here, 
minimizing the bounces leads to a unique solution. Both sets (a) and (b)
have regions where all bounce probabilities are zero, as shown in
Fig.~\ref{nobounce}(a). The ``algorithmic phase'' diagram for the full
minimum-bounce solution has four different regions, as shown in
Fig.~\ref{nobounce}(b). They correspond to different analytical
forms of the solution. 
All the path-weights in these regions are listed in Table 1. As discussed
above, the actual probabilities are simply obtained by normalizing with the 
matrix elements $W_k$ of the corresponding equations in (\ref{eqsets}).

The solutions in the regions A-D in Fig.~\ref{nobounce}(b) are continuous
across the boundaries. In particular, it can be noted that when the
isotropic Heisenberg point, $\Delta=1,h_b=0$, is approached, and the 
minimum value $\epsilon_{\rm min}$ of the constant $\epsilon$ is used,
the only surviving vertex process is the switch-and-reverse, corresponding
to the weights $a_3$, $b_3$ in Fig.~\ref{dpaths}. This is exactly the
process used in the deterministic update illustrated in Fig.~\ref{sseloop}.
Hence, the general directed loops indeed smoothly reduce to these very 
special symmetric loops, that were constructed in a different manner by 
using the fact that the diagonal and off-diagonal matrix elements have 
the same values $0,1/2$ at the isotropic point.

It is interesting to note that the constant $\epsilon$ appears only in the 
weights $a_1$, $b_1$, corresponding to the ``continue-straight'' process
in Fig.~\ref{dpaths}. One can always choose $\epsilon=\epsilon_{\rm min}$
(which has the advantage that it minimizes the average expansion-order 
$\langle n\rangle$ and the cut-off $M$), but in some cases when 
$\epsilon_{\rm min}=0$ a non-zero $\epsilon$ can lead to shorter 
autocorrelation times \cite{dloop}. In a world-line 
formulation of the directed-loop update \cite{dloop}, the constant 
$\epsilon$ does not appear at all, but apart from this the path-weights 
are the same as in Table 1. When taking the continuum limit of the 
time-discretized world-lines, it has been shown \cite{dloop} that the
vertex-probabilities reduce exactly to those obtained before 
\cite{evertz2,beard} on the line $h=0$, $\Delta \le 1$. This smooth reduction
of the directed loops to the conventional self-avoiding loops shows that this 
scheme is a natural generalization of the loop concept when the requirement of
self-avoidance is relaxed. In the earlier generalized loop algorithms, 
i.e., the worm algorithm \cite{worm} and the SSE operator-loop algorithm with
the simple heat-bath probabilities \cite{sseloop} (which also correspond
to a solution of the directed-loop equations), the bounce probability 
does not vanish as $h \to 0$, and hence any formal relationship between
the generalized and conventional loop algorithms was unclear. The 
relationships between the various loop-type methods are also reviewed 
elsewhere in this Volume \cite{troyer2}.

Practical implementation details of the directed-loop algorithm have not
been discussed here; they have been outlined in Ref.~\cite{dloop}. Some
example-programs are also available on-line \cite{online}.

\begin{table}
\begin{tabular}{llllllll}
\hline
  & \tablehead{1}{l}{b}{$\epsilon_{\rm min}$ \\ ~} 
  & \tablehead{1}{l}{b}{$a_1$ \\ $b_1$} 
  & \tablehead{1}{l}{b}{$a_2$ \\ $b_2$} 
  & \tablehead{1}{l}{b}{$a_3$ \\ $b_3$} 
  & \tablehead{1}{l}{b}{$a_4$ \\ $b_4$} 
  & \tablehead{1}{l}{b}{$a_5$ \\ $b_5$} 
  & \tablehead{1}{l}{b}{$a_6$ \\ $b_6$} \\
\hline
$A$~~~ & $\Delta^- - f$~~
    & $\epsilon + f - \Delta^-$~~
    & $\Delta^- -f$~~
    & $\Delta^+ +f$~~
    & $0$~~
    & $0$~~ 
    & $0$~~ \\
~   & ~ 
    & $\epsilon +3f - \Delta^-$~~ 
    & $\Delta^- +f$~~
    & $\Delta^+ -f$~~
    & $0$~~
    & $0$~~ 
    & $0$~~  \\
\hline
$B$~~~ & $0$~~
    & $\epsilon$~~
    & $0$~~
    & $1/2$~~ 
    & $0$~~
    & $2(f-\Delta^-)$~~
    & $0$~~ \\
~   & ~ 
    & $\epsilon +3f - \Delta^-$~~ 
    & $\Delta^- +f$~~
    & $\Delta^+ -f$~~
    & $0$~~
    & $0$~~ 
    & $0$~~  \\
\hline
$C$~~~ & $0$~~ 
    & $\epsilon$~~
    & $0$~~
    & $1/2$~~
    & $0$~~
    & $2(f-\Delta^-)$~~ 
    & $0$~~ \\
~   & ~ 
    & $\epsilon +4f$~~
    & $0$~~
    & $1/2$~~ 
    & $0$~~
    & $-2(\Delta^- +f)$~~
    & $0$~~  \\
\hline
$D$~~~ & $0$~~
    & $\epsilon$~~
    & $0$~~
    & $1/2$~~ 
    & $0$~~
    & $2(f-\Delta^-)$~~
    & $0$~~ \\
~   & ~ 
    & $\epsilon +2f +\Delta/2$~~
    & $1/2$~~
    & $0$~~
    & $2(f-\Delta^+)$~~
    & $0$~~ 
    & $0$~~  \\
\hline
\end{tabular}
\caption{Vertex-path weights and the minimum value of the constant $\epsilon$ 
in the different regions of Fig.~\ref{nobounce}(b). A short-hand notation 
$\Delta^\pm = (1 \pm \Delta)/4$, $f=h_b/2$ is used.}
\label{tab1}
\end{table}

\section{Applications and generalizations}

The directed-loop algorithm has already been applied to several models in
addition to the $S=1/2$ antiferromagnetic XXZ model discussed here and in 
Ref.~\cite{dloop} (the ferromagnetic case can be treated in a very similar
manner \cite{dloop}). An identical algorithm in the continuous-time 
world-line representation has been applied in a large-scale study of the 
weakly anisotropic ($\Delta > 1,h=0$) system \cite{cuccoli}. 
Extensions to higher spins and softcore boson models have been explored 
by several groups 
\cite{sara,harada,syljuasen,alet1}. Four-spin interactions have also
been considered \cite{jkmodel}. An application to the 1D extended 
Hubbard model has produced high-precision results \cite{jcomment} for 
larger systems sizes than what was practically feasible with previous 
methods. A different type of directed loops have been developed for 
quantum-rotor models \cite{alet2}. 

In general, for $S > 1/2$ and softcore bosons, minimization of the bounce 
probability does not lead to a unique solution of the directed-loop
equations \cite{harada}, 
and therefore some other constraints have to be applied as well. 
It has also been pointed out that the minimization of the bounce is not 
necessarily the optimal strategy \cite{alet1} (which was also anticipated 
not to be strictly true from the outset \cite{dloop}). Also, for the 
Heisenberg model with $S>1$, in order to eliminate the bounces completely 
one has to assign multiplicative weights $\not=1$ also to the discontinuities
(sources) that exist while the loop is being constructed \cite{alet1}.

It appears that in most cases it is relatively easy to find low-bounce 
solutions that work well in practice, but it remains a challenging problem 
to find a scheme for automatically generating the optimal 
vertex-probabilities, i.e., without carrying out time-consuming
tests of actual simulation programs. The strength of the directed-loop 
scheme is that it provides a well-defined, general mathematical framework 
for this pursuit.


\begin{theacknowledgments}
We would like to thank F. Alet, P. Henelius, N. Kawashima, M. Troyer, 
and S. Wessel for stimulating discussions. This work was supported 
by the Academy of Finland and by a Nordic network project on Strongly 
Correlated Electrons at NORDITA.
\end{theacknowledgments}


\bibliographystyle{aipproc}   



\end{document}